\documentclass[times,twocolumn,final,authoryear]{elsarticle}

\usepackage{pseudoprletters}
\usepackage{framed,multirow}

\usepackage{amssymb}
\usepackage{latexsym}

\usepackage{url}
\usepackage{xcolor}
\definecolor{newcolor}{rgb}{.8,.349,.1}

\journal{Pattern Recognition Letters}

\begin{document}

\ifpreprint
  \setcounter{page}{1}
\else
  \setcounter{page}{1}
\fi

\begin{frontmatter}

\title{The Fuzzy ROC}

\author[1]{Giovanni \snm{Parmigiani}\corref{cor1}} 
\cortext[cor1]{Corresponding author: Giovanni Parmigiani}
\ead{gp@jimmy.harvard.edu}

\address[1]{Dana Farber Cancer Institute, 450 Brookline Avenue, Boston
  02115, U.S.A. \\ and Harvard T.H. Chan School of Public Health, 677
  Huntington Avenue, Boston 02115, U.S.A. }

%\received{1 May 2013}
%\finalform{10 May 2013}
%\accepted{13 May 2013}
%\availableonline{15 May 2013}
%\communicated{S. Sarkar}

\begin{abstract}
The fuzzy ROC extends Receiver Operating Curve (ROC) visualization to the situation where some data points, falling in an indeterminacy region, are not classified. It addresses two challenges: definition of sensitivity and specificity bounds under indeterminacy; and visual summarization of the large number of possibilities arising from different choices of indeterminacy zones. 
\end{abstract}

\begin{keyword}
\KWD Receiver Operating Curves (ROC) \sep Indeterminacy in classification
 %Keyword1\sep Keyword2\sep Keyword3

\end{keyword}

\end{frontmatter}

%\linenumbers

\section{Introduction}
\label{intro}

Receiver Operating Curves (ROC) help with the visual assessment of the
performance of classifiers. \cite{FAWCETT2006861} reviews the field
and points out that ``ROC graphs are commonly used in medical decision
making, and in recent years have been used increasingly in machine
learning and data mining research''.  

I consider here the basic case of binary classification using a
continuous score, such as a classification probability,
or a quantitative biomarker. Traditionally, classification is simply
implemented by a cutoff dichotomizing the score. In more recent
applications, classification may includes an
intermediate area of indeterminacy, which I will call {\it gray zone}.  

For a famous example, \cite{Parker:2009gxa} present the PAM50 risk
predictor of breast cancers, which provides a continuous risk
score. In clinical applications, this score is most often split into
three categories: low, intermediate and high. Women in the low and
high categories are directed to specific clinical strategies. Women in
the intermediate category are considered on a case by case basis by their
clinicians. From an algorithmic standpoint, the intermediate group is
not classified. Similarly, machine learning algorithms for
classification of pathology and 
radiology images may allow for certain areas
to be routed to further human examination. In these cases
indeterminacy helps with practical implementation, by handling safe
cases algorithmically and complex ones by human intervention.

Here I describe an algorithm for visualizing bounds on
sensitivity/specificity pairs, for short {\it fuzzy
ROC}, to assess the performance range of classifiers allowing for a
region of indeterminacy, or gray zone. I try to address two
challenges. The first is the definition of sensitivity and
specificity bound when there is indeterminacy. The second is the visual summarization of the large number of possibilities arising from different choices of gray zones. 
\section{Algorithm}

Consider a validation study of $n$ labeled subjects, with scores
$x_i$, $1=1,\ldots,n$.  
Without loss, let the first $n_0>0$ subjects have label $0$ and the rest have
label $1$. Also, low levels of the score are taken to predict class $0$. The
proportion of $1$'s in the target population is $\pi$, and may differ
from the validation study proportion $n_1 / n$, for example if the design
of the validation study is a case-control.  

A gray zone is defined by the interval $(c_L,c_H)$. The extremes are
the lower and higher cutoff. Cases with score below $c_L$ are
classified as $0$'s. Cases above $c_H$ are classified as $1$'s. The
rest remain unclassified.

Users of fuzzy ROC need to specify a maximum tolerated percentage of unclassified cases, $\gamma$. 
Let $g_j$ be the number of class $j$ points falling in the gray zone.
A gray zone $(c_L,c_H)$ satisfies the $\gamma$-constraint if the proportion of cases in the gray zone is less than $\gamma$, that is if $(g_0+g_1)/n < \gamma$. A gray zone $(c_L,c_H)$ satisfies the target population $\gamma$-constraint if $((1-\pi) g_0+\pi g_1)/n < \gamma$.

The fuzzy ROC algorithm is a model-free visualization. The basic
building blocks are bounds on the cumulative frequencies associated with a given
gray zone $(c_L,c_H)$.  

First, the most favorable bound on these frequencies is calculated
assuming perfect discrimination within the gray zone. Imagine an
oracle would take care of the points in the gray zone on behalf of the
classifier, by moving them to the extremes of the gray zone so that
they can be classified correctly. Formally, define the starred scores as follows: 
\[
\begin{array}{cccrcl}
\mbox{if}  & x_i \not\in (c_L,c_H)  &  \mbox{then} & x_i^* &=& x_i \\
\mbox{if}  & i \leq n_0, x_i \in (c_L,c_H)  & \mbox{then} & x_i^* &=& c_L  \\
\mbox{if}  & i > n_0, x_i \in (c_L,c_H)   &   \mbox{then} & x_i^* &=& c_H.
\end{array}
\]
Let $I_A$ be the indicator of the set $A$, and define the cumulative frequencies: 
\begin{eqnarray}
F_0^* (c_L,c_U) & = & \sum_{i=1}^{n_0} I_{ x_i^* < (c_L+c_U)/2 } \\
F_1^* (c_L,c_U) & = & \sum_{i=n_0+1}^{n} I_{ x_i^* < (c_L+c_U)/2 }.
\end{eqnarray}

Conversely, the least favorable frequencies are constructed considering the worst case scenario for the points within the gray zone. Imagine now that a saboteur may be in charge of the points in the gray zone, by moving them to extremes of the gray zone, so that they are all classified incorrectly. This would result in the "daggered" scores, defined as: 
\[
\begin{array}{cccc}
\mbox{if}  & x_i \not\in (c_L,c_H)  &  \mbox{then} & x_i^{\dagger} = x_i \\
\mbox{if}  & i \leq n_0, x_i \in (c_L,c_H)  & \mbox{then} & x_i^{\dagger} = c_H  \\
\mbox{if}  & i > n_0, x_i \in (c_L,c_H)   &   \mbox{then} & x_i^{\dagger} = c_L.
\end{array}
\]
Now define the cumulative frequencies: 
\begin{eqnarray}
F_0^{\dagger} (c_L,c_U) & = & \sum_{i=1}^{n_0} I_{ x_i^* < (c_L+c_U)/2 } \\
F_1^{\dagger} (c_L,c_U) & = & \sum_{i=n_0+1}^{n} I_{ x_i^* < (c_L+c_U)/2 }.
\end{eqnarray}

We can form a large number of starred and daggered pairs of
cumulative frequencies satisfying the $\gamma$-constraint.
The fuzzy ROC algorithm simplifies the visualization of these pairs by grouping them, and selecting a single higher and lower limit within each group, as follows.  

Consider the $r$ observed unique ranked values of the biomarker $x_{(1)},
\ldots x_{(r)}$. These points will constitute the set of possible values for the extremes $(c_L,c_H)$ of the gray zone. Now define the midpoints between two
consecutive values as $c_j = ( x_{(r-1)} + x_{(r)} ) / 2$ for $j = 2,
\ldots, r$. 
For each $c_j$, consider the set of 
$(c_L,c_H)$ pairs built by first adding the two neighboring observed points on either side, then the next two and so forth. This process continues as long as the gray zone satisfies the $\gamma$-constraint. If one of the extremes of the distribution is reached, the process continues on the other side. Among the resulting intervals, the fuzzy ROC chooses the "best" for visualization, defined as follows.
For each $(c_L,c_H)$, it eliminates the cases in the gray zone and then computes the AUC curves on the classified cases only. 
The $(c_L,c_H)$ pair maximizing the AUC so defined is
$(c_L^*(c_j),c_H^*(c_j))$.
The generating $c_j$ is not necessarily the midpoint of this interval,
but will be contained in it.
If multiple gray zones are tied in this maximization, the algorithm
minimizes gray zone size among optima. 
In this way, gray zones are not used in regions where discrimination is not
helped by not classifying cases.

Then, the upper limits are defined by the set of points
\begin{equation}
\left( 1-F_1^* (c_L^*(c_j),c_H^*(c_j)), 1-F_0^* (c_L^*(c_j),c_H^*(c_j)) \right)
\end{equation}
as $c_j$ varies. 
Conversely, the lower limits are defined by the set of points
\begin{equation}
\left( 1-F_1^{\dagger} (c_L^*(c_j),c_H^*(c_j)), 
  1-F_0^{\dagger} (c_L^*(c_j),c_H^*(c_j)) \right).
\end{equation}
for $j = 2, \ldots, r$.  To implement, define the degenerate gray zones
$(x_{(i)}, x_{(i)})$ and $(x_{(i)}, x_{(i+1)})$ as the empty set.

Fix $y$ to be either $0$ or $1$.  The sequences defined by
$F_y^*(c_L^*(c_j),c_H^*(c_j))$ and
$F_y^{\dagger}(c_L^*(c_j),c_H^*(c_j))$ as $j$ varies in $2, \ldots, r$
do not necessarily define proper cumulative distributions, as they would in a standard ROC analysis. Rather
the intent is to provide bounds to the sensitivity / specificity pairs
available over a range of possible gray area strategies.

Starred and daggered curves are calculated using both classified and unclassified
samples. The exclusion of the unclassified samples only affects the
calculation of $(c_L^*(c_j),c_H^*(c_j))$.

I explored an alternative implementation where the lower and upper
limit of the gray area are used in turn to index the AUC optimization,
instead of the midpoints.  Upper and lower limits can produce markedly
different results. Bounds are less stable than the midpoints when
sample sizes are small.  Nonetheless, this strategy provides a
different view of the overlap in the tails, and may turn out to be useful in
some applications.

\section{Illustration}

To illustrate the application and interpretation of the fuzzy ROC, I
consider a gene expression biomarker for the prediction of suboptimal
(class~$0$) versus optimal (class~$1$) surgical debulking in ovarian
cancer patients. Data are available from the {\tt CuratedOvarianData}
Bioconductor package  by \cite{Ganzfried2013}. Clinical and biological
background can be found in \cite{Riester2014}. The specific biomarker
presented here reflects the transcriptional level of the gene ZNF544,
as measured using an Agilent microarray by \cite{Yoshihara1374}. 

\begin{figure}[t]
\includegraphics[width=.49\textwidth]{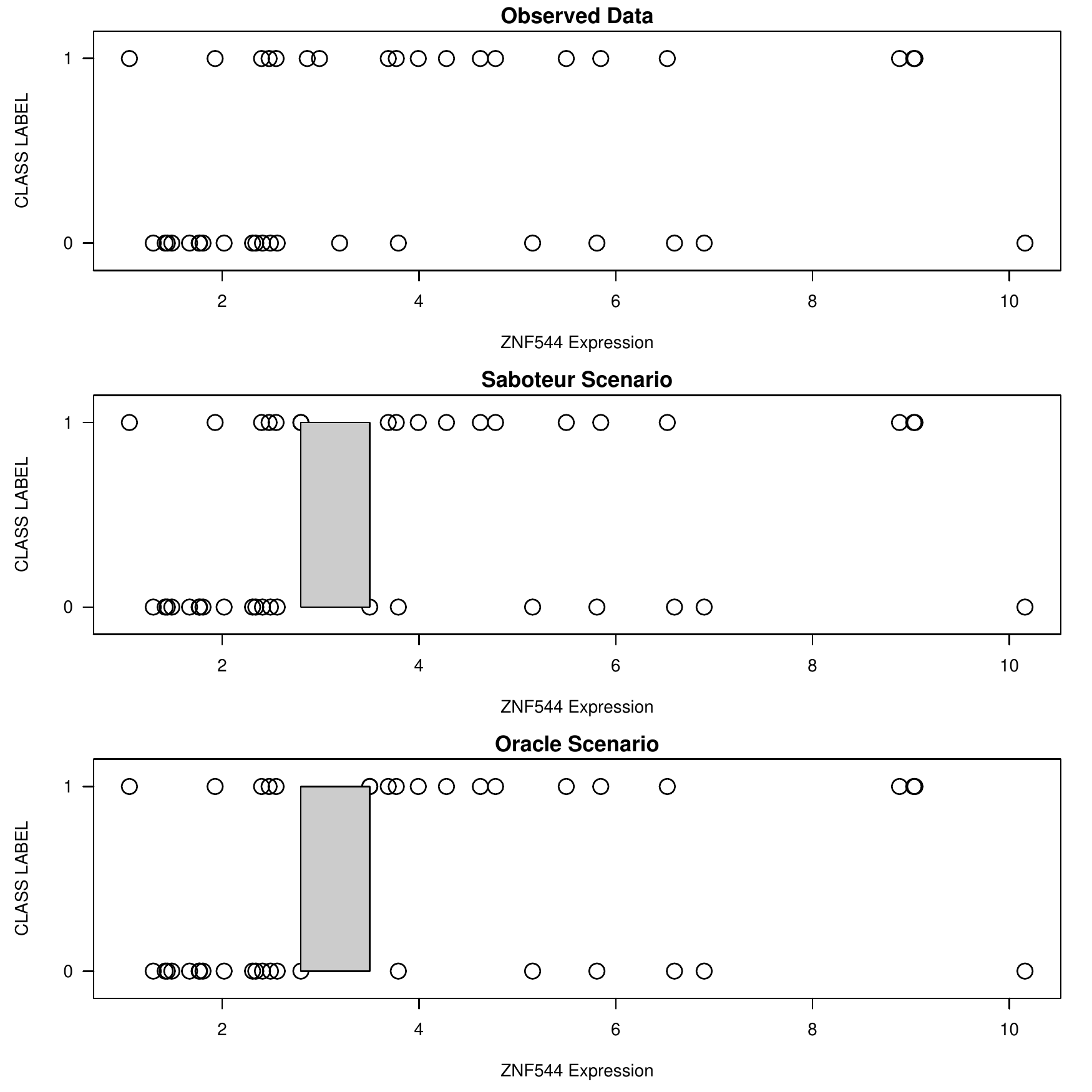}
\caption{Dotplots of biomarker levels by class as observed (top) and
  in the
  hypothetical scenarios used in the construction of the fuzzy ROC plot. The gray
  zone is $(2.8,3.5)$. In the
  ``oracle'' scenario, class 1 points in the gray zone are moved to
  the upper limit $3.5$ while the class 0 points in the gray zone are
  moved to $2.8$. The reverse is true in the ``saboteur'' scenario. \label{fig:dotplot}}
\end{figure}

Figure~\ref{fig:dotplot} shows the observed biomarker levels by
class. Higher level of expression are generally associated with
optimal debulking (class 1). 
Figure~\ref{fig:dotplot} also illustrates the type of
hypothetical scenarios that enter as building block in the
construction of the fuzzy ROC, to visually represent the definitions of
$x^*$ and $x^\dagger$.

\begin{figure}[t!]
\includegraphics[width=.49\textwidth]{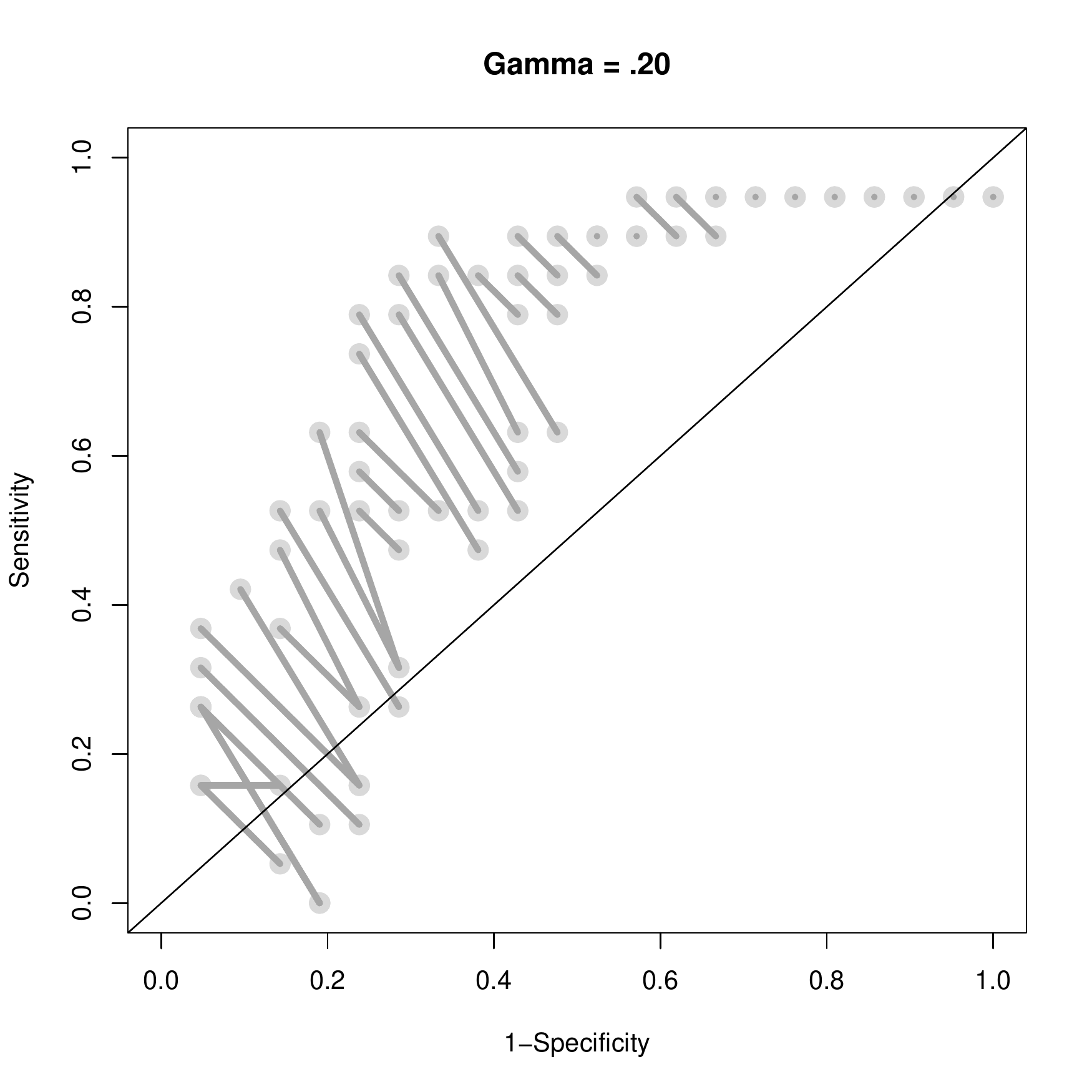}\\
\includegraphics[width=.49\textwidth]{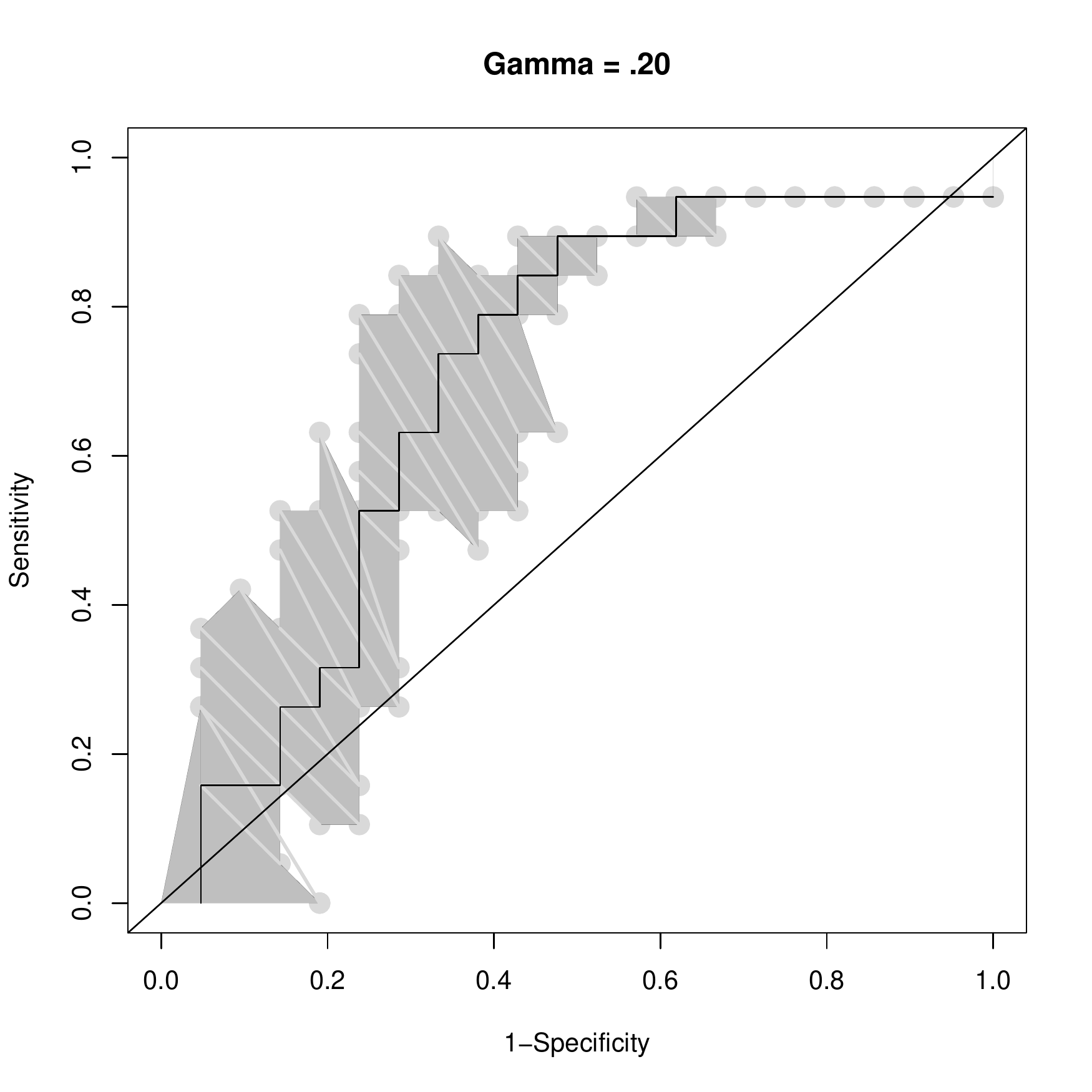}
\caption{Fuzzy ROC displays at maximum tolerated percentage of
  unclassified cases, $\gamma$, of~$.2$. The top panel shows segments
  connecting starred and daggered points corresponding to the same
  $c_j$. The segments collapse to a point when the optimal gray area for
  the corresponding $c_j$ is empty. The bottom panel shows, in
  addition, the area between the two 
  curves defined by connecting the starred and daggered points. 
The thinner line corresponds to the standard ROC curve. \label{fig:fuzzyROC}}
\end{figure}

Each of hypothetical scenarios in Figure~\ref{fig:dotplot} enter the
optimization used to find the $c_H^*(c_j)$'s. These in turn are used
to form the starred and daggered sensitvity and specificity bounds. 
Figure~\ref{fig:fuzzyROC} shows segments
  connecting starred and daggered points corresponding to the two
  bounds associated with the same
  $c_j$. These can be used to explore potential gray area strategies. 
Say one is interested in a classifier with approximately 80\% specificity and 70\%
sensitivity. ZNF544 does not reach this performance. The upper points
inform us that if one were allowed to pass
20\% of suitably chosen observations to the oracle, than ZNF544 could
reach close to the
desired sensitivity/specificity trade-off. It also informs us that if the
same observations were passed to the saboteur, the sensitivity and
specificity would drop close to the diagonal line of no discrimination.

Figure~\ref{fig:fuzzyROC} also shows, in the right panel, the region
defined by the starred points as the upper limit, and by the daggered points as the lower limit. Points
within the region are not easily interpretable in terms of the
optimization of the previous section. The shading is purely a visual aid.

\begin{figure}[t!]
\begin{center}
\includegraphics[width=.23\textwidth]{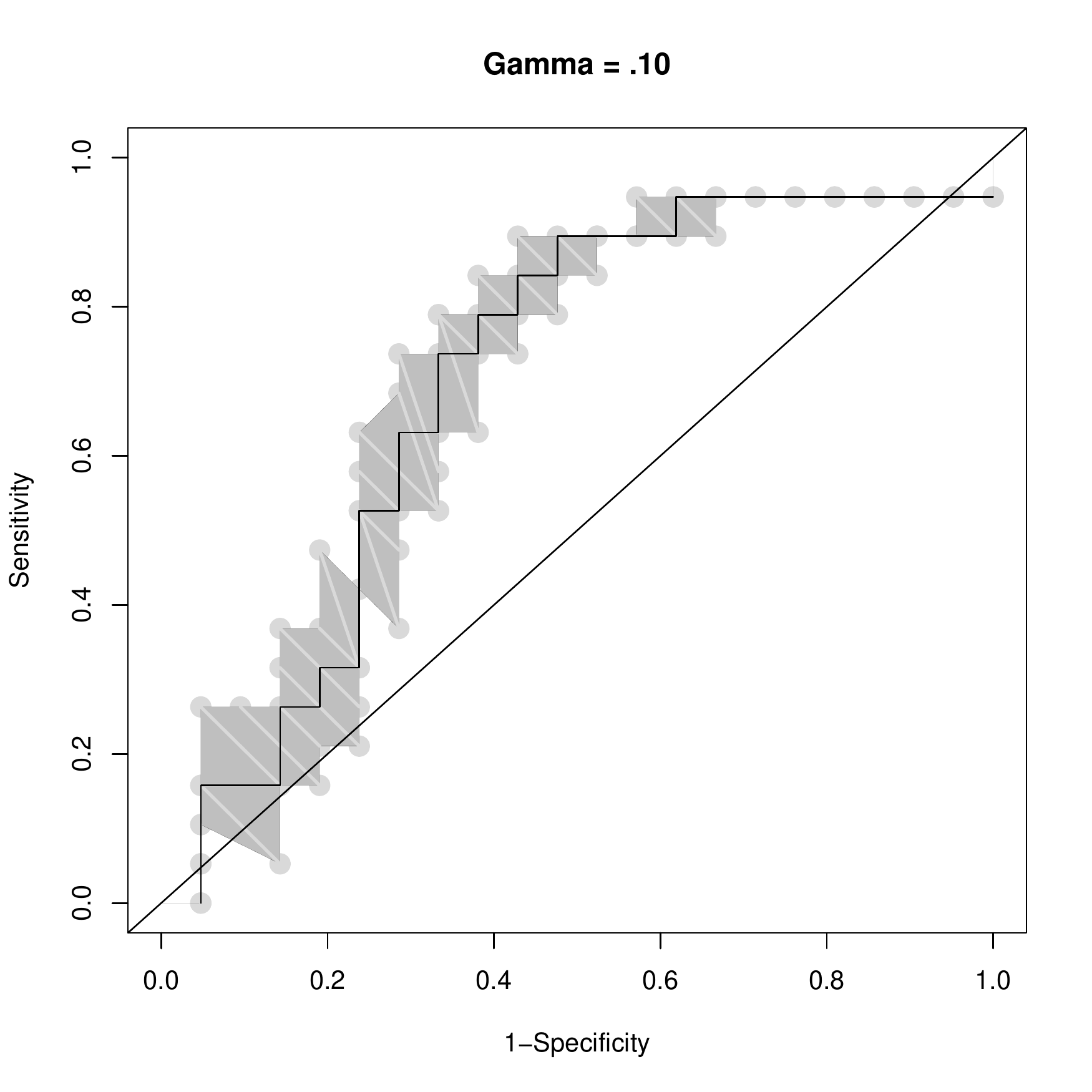}
\includegraphics[width=.23\textwidth]{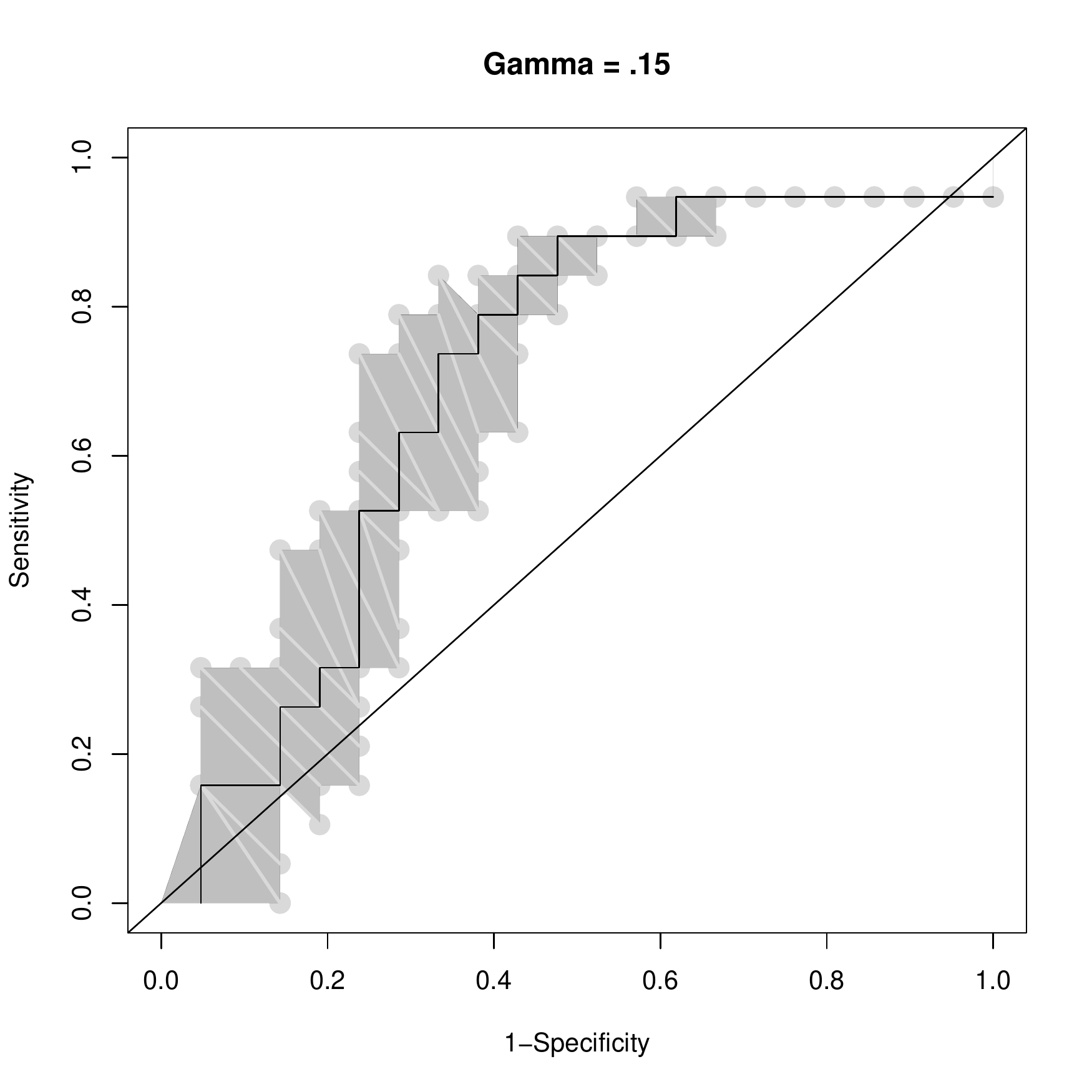}\\
\includegraphics[width=.23\textwidth]{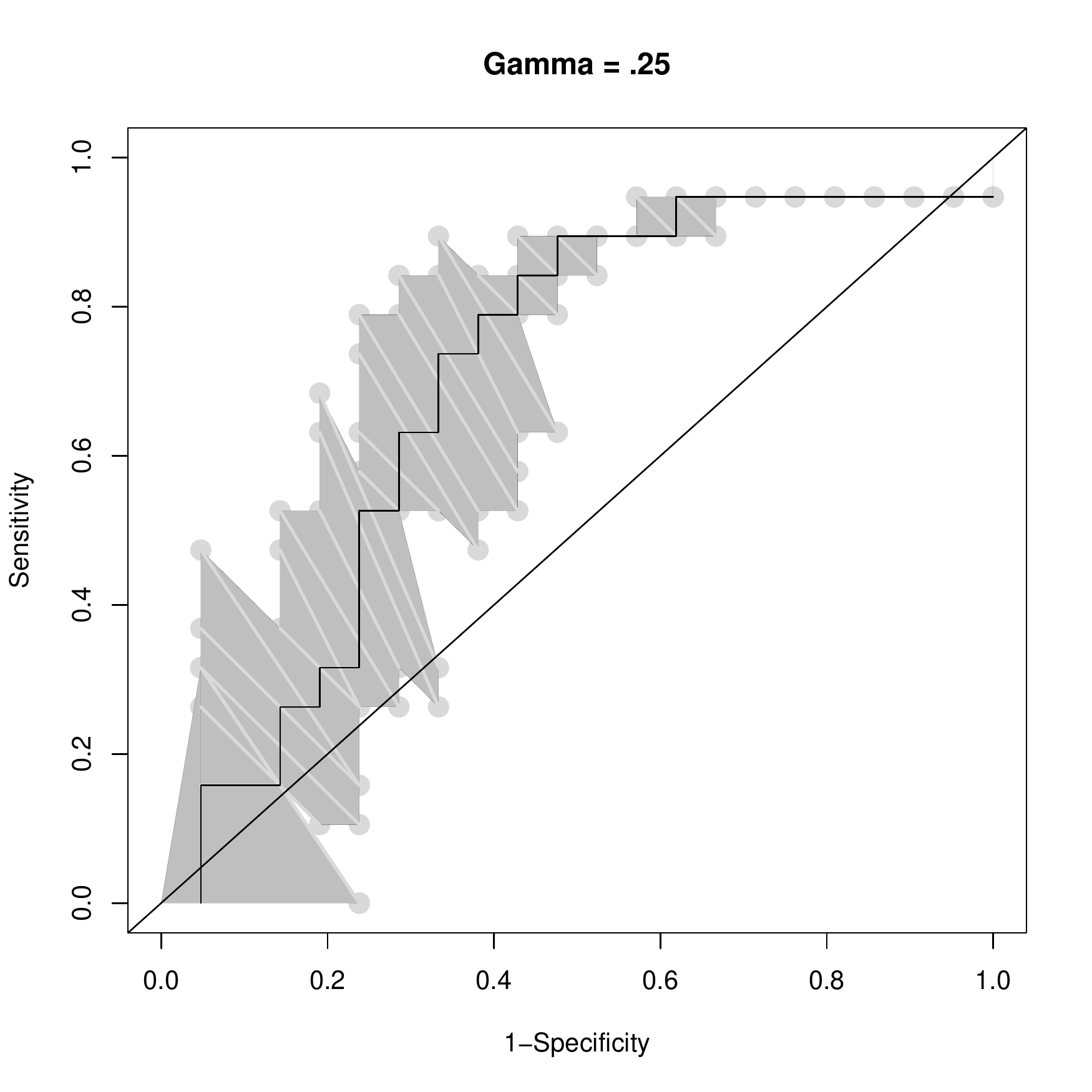}
\includegraphics[width=.23\textwidth]{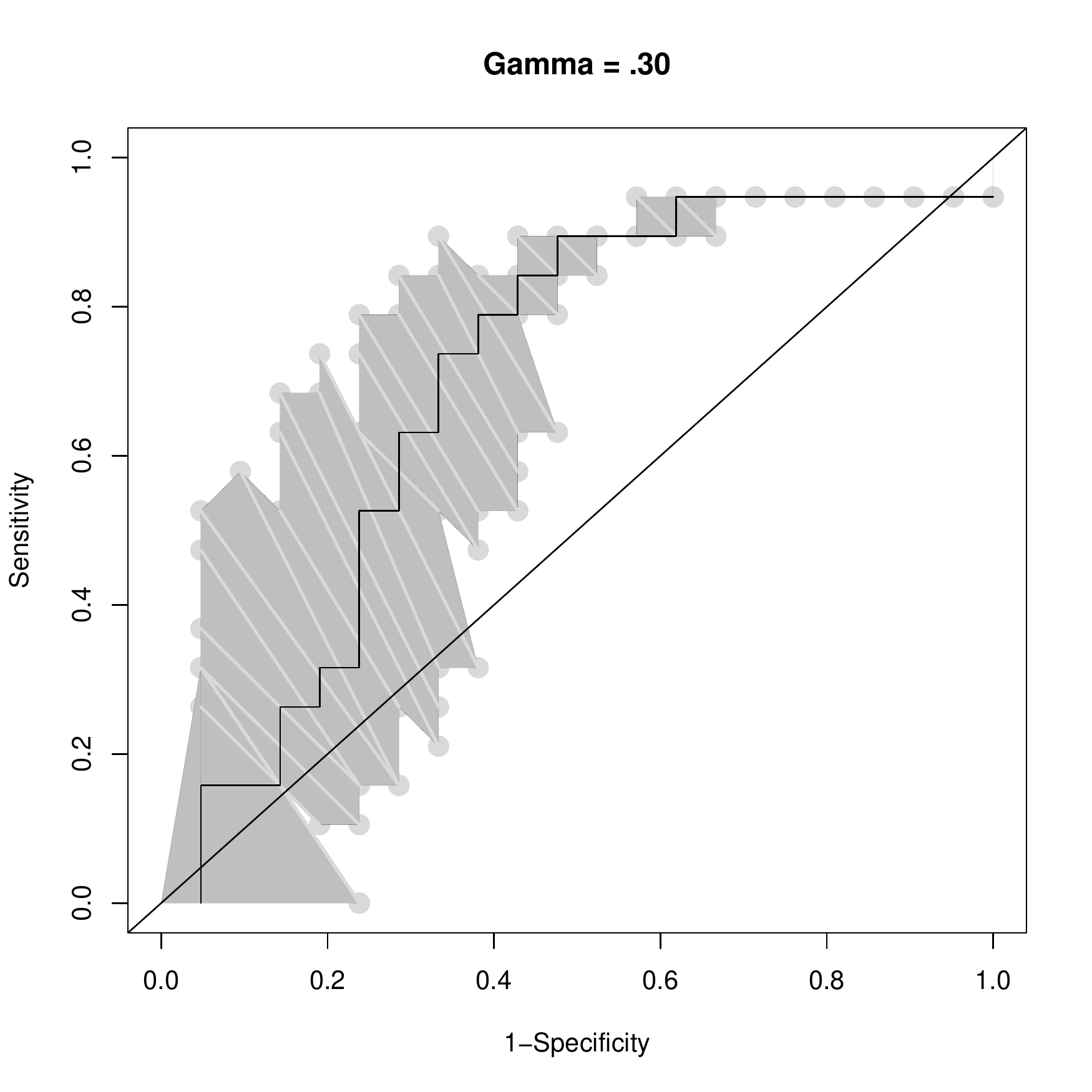}
\end{center}
\caption{Fuzzy ROC displays for ZNF544 at maximum tolerated percentage of
  unclassified cases, $\gamma$, of $.1$ (top left) $.15$ (top right)
  $.25$ (bottom left) and $.30$
  (bottom right.) The thinner line corresponds to the standard ROC curve. \label{fig:fuzzyROCgammas}}
\end{figure}

Figure~\ref{fig:fuzzyROCgammas} shows fuzzy ROC visualizations
corresponding to four additional choices of $\gamma$. 

Figure~\ref{fig:fuzzyROC} also illustrates that the region defined by
the upper and lower limits in the fuzzy ROC algorithm is not
necessarily convex.

If $\gamma=0$ the fuzzy ROC region collapses to the standard ROC line,
also drawn in Figures~\ref{fig:fuzzyROC} and~\ref{fig:fuzzyROCgammas}. 

\begin{figure}[t]
\includegraphics[width=.23\textwidth]{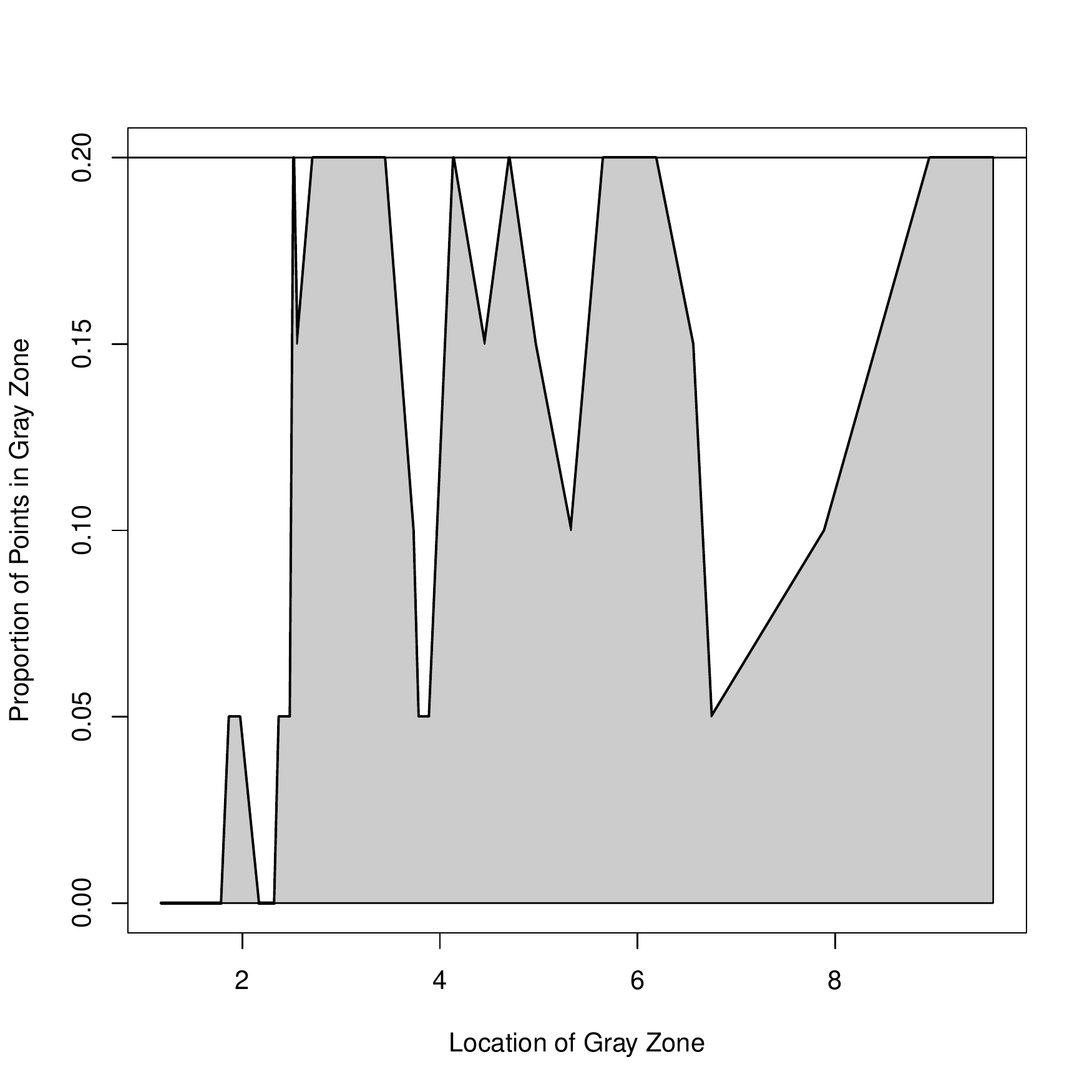}
\includegraphics[width=.23\textwidth]{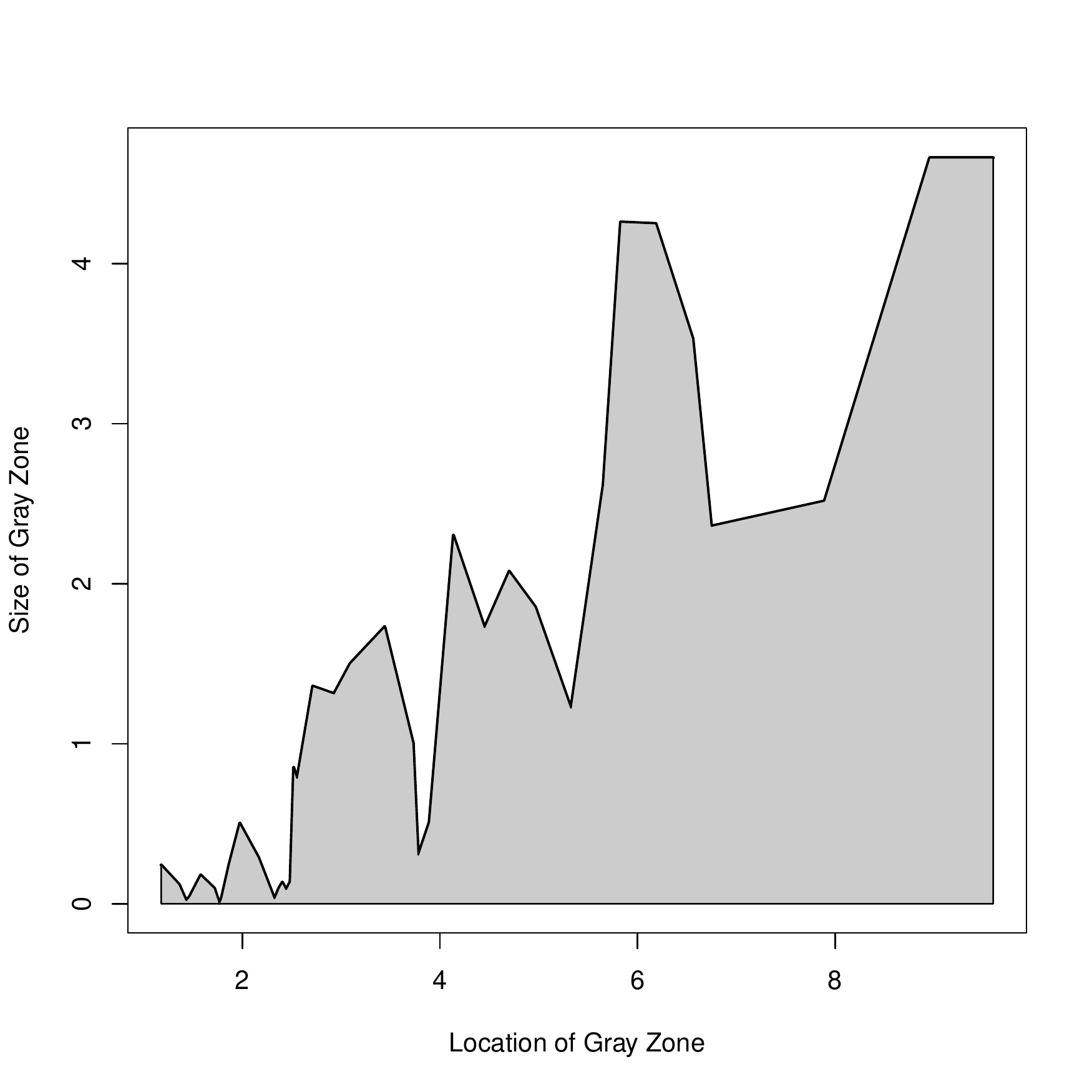}
\caption{Proportion of points falling in the gray zone (left) and width
  of the gray zone in the biomarker scale (right) as a function of
  $c_j$ at $\gamma=.2$. \label{fig:grayZone}}
\end{figure}

In regions where the two class-specific distributions have little
overlap, say left of $2$, there can be little or no advantage in
allowing for a gray zone. Conversely, where the density of biomarker
points in the two classes is similar, a gray zone has
the potential to improve the practical implementation of the
biomarker. Figure~\ref{fig:grayZone} depicts this trade-off by
elucidating where in the biomarker range the gray area is useful. Only
in a narrow range of values does the fuzzy ROC algorithm needs to make
full use of the 20\% of data points allowed for the gray zone (top
panel).

\begin{figure}[t!]
\includegraphics[width=.23\textwidth]{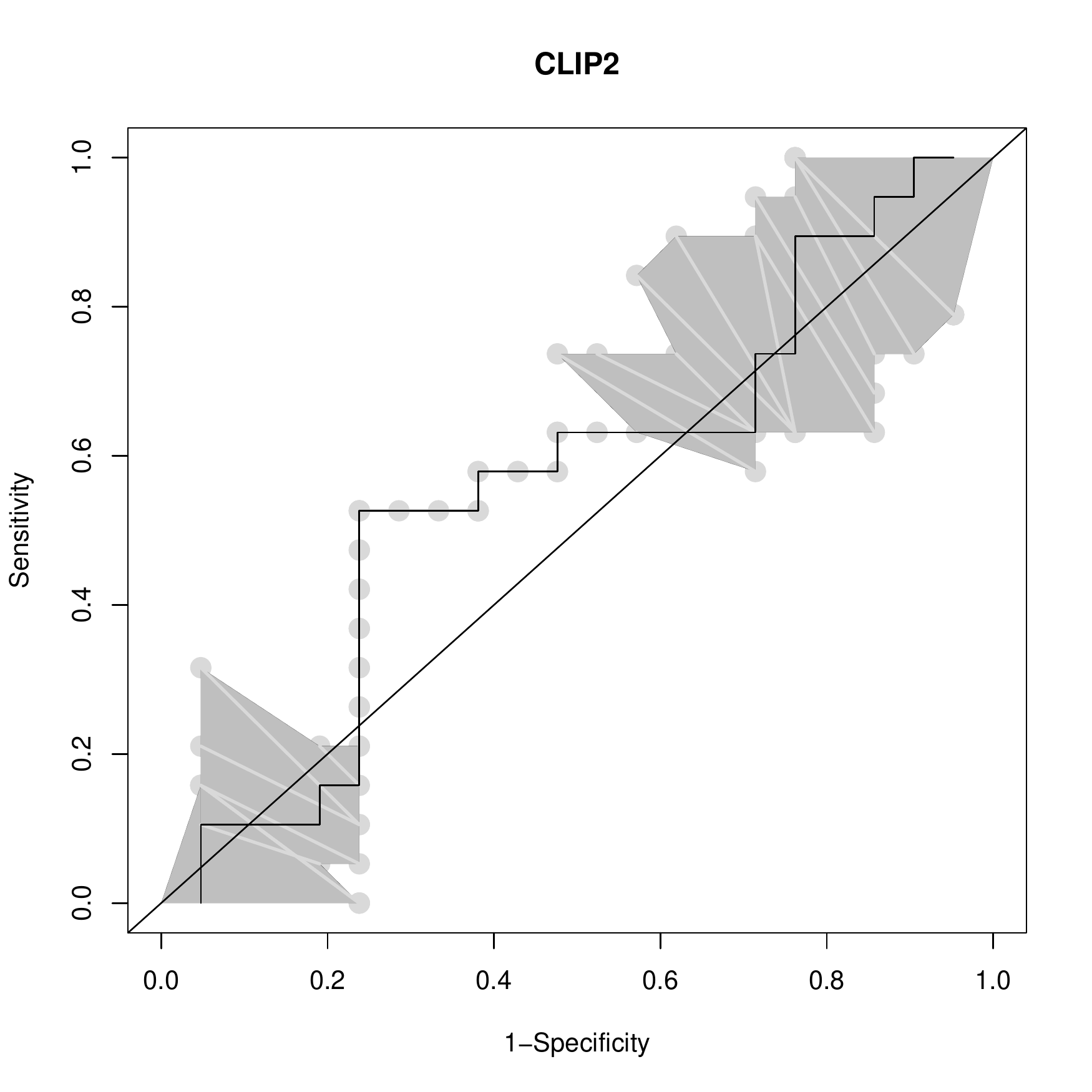}
\includegraphics[width=.23\textwidth]{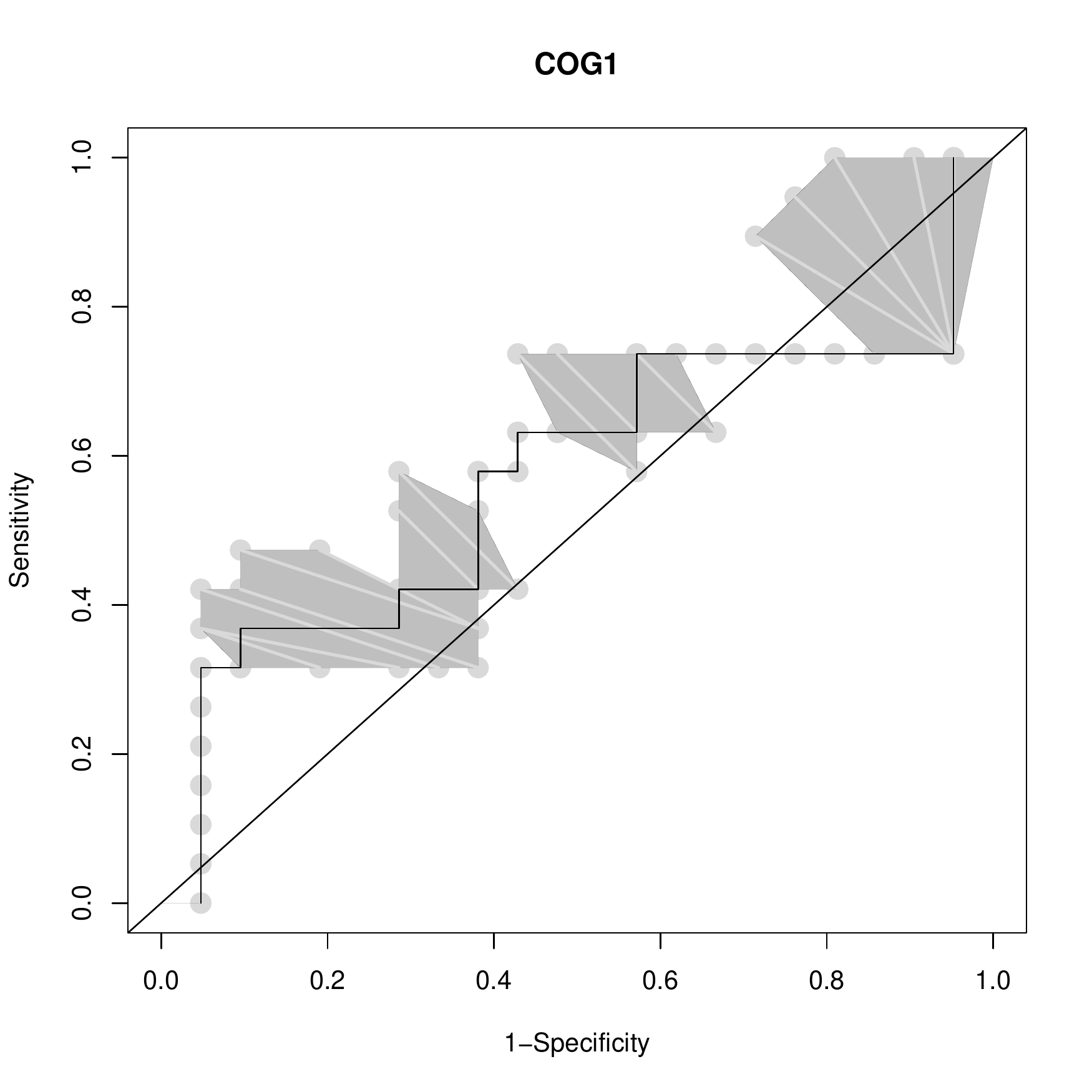} \\
\includegraphics[width=.23\textwidth]{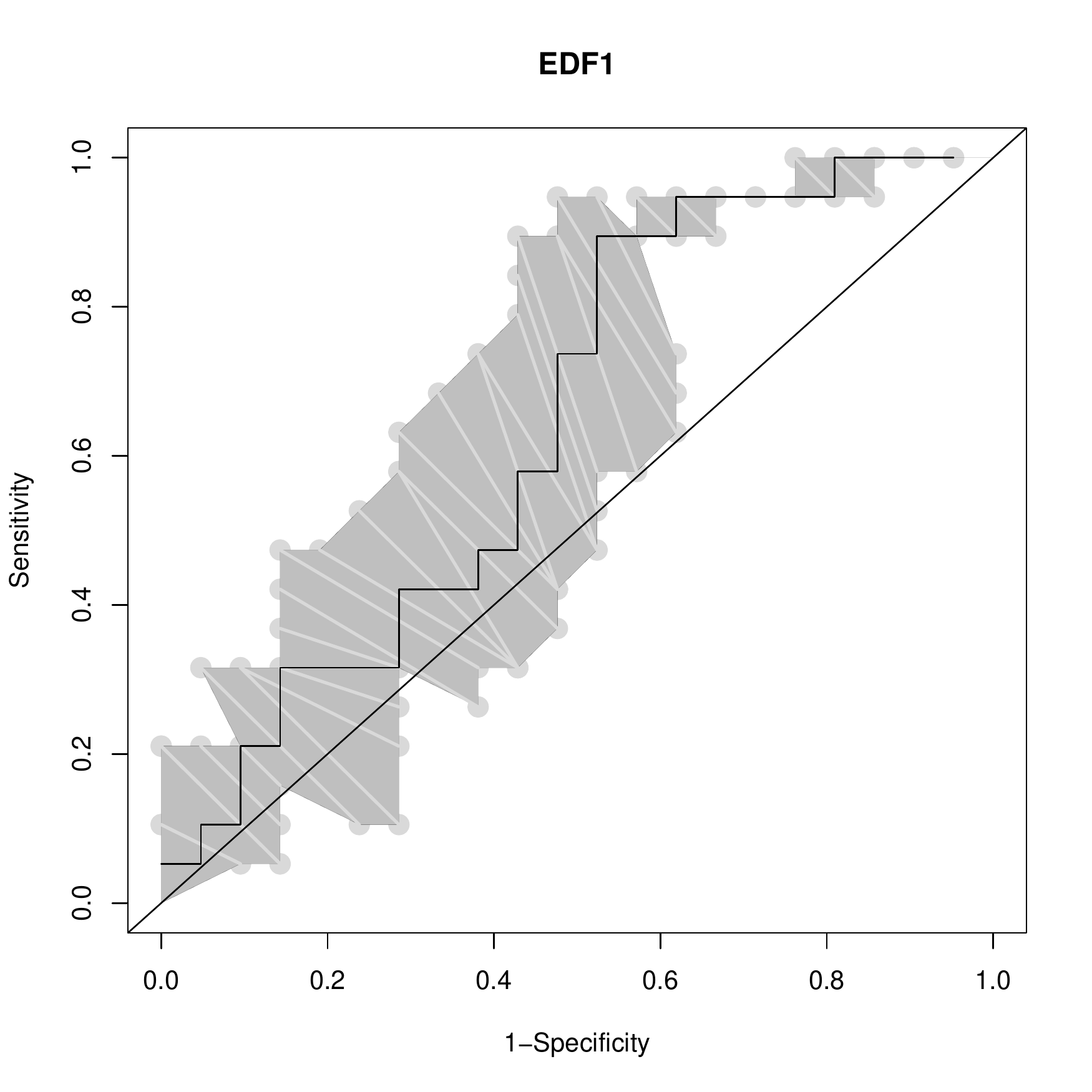}
\includegraphics[width=.23\textwidth]{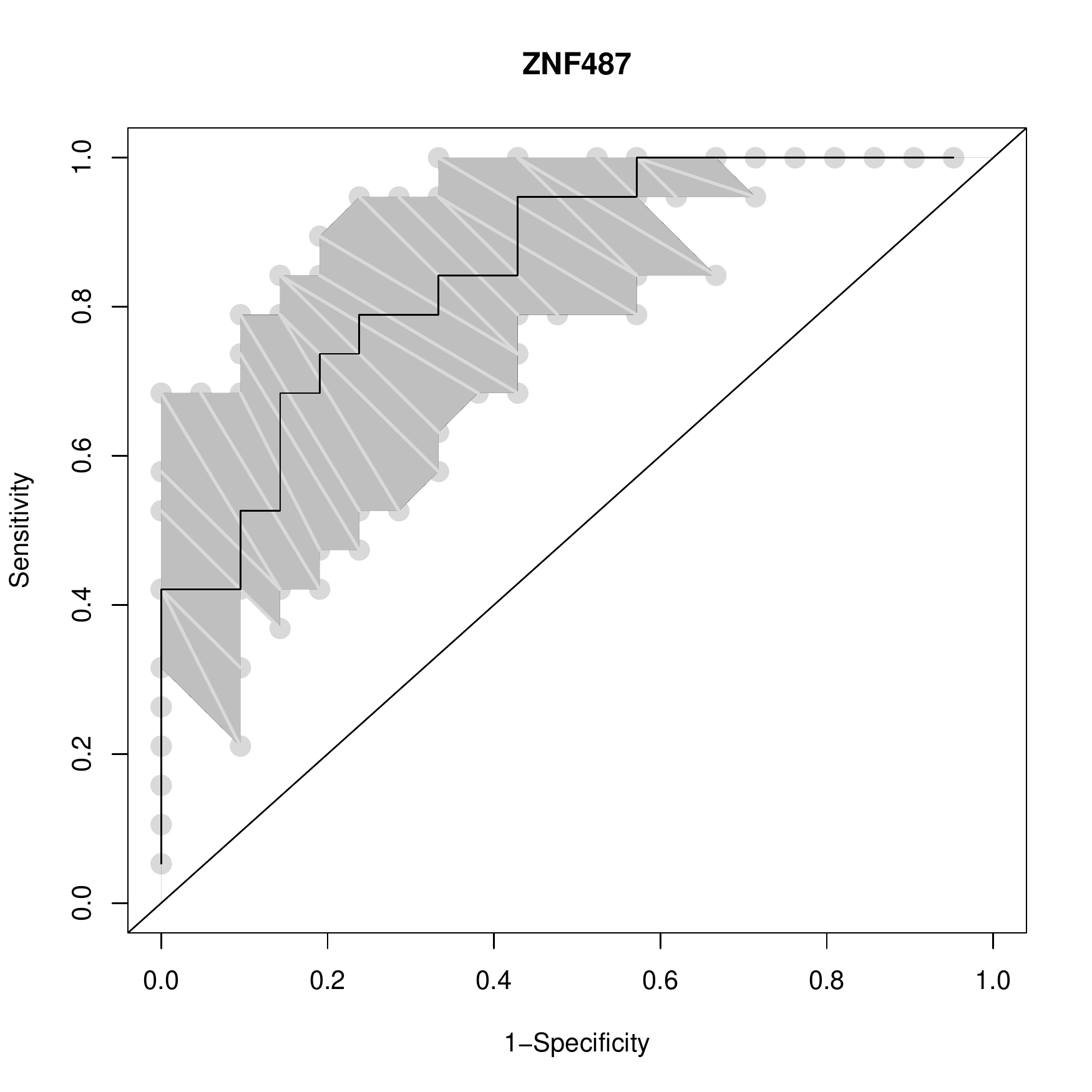}
\caption{Fuzzy ROC displays at maximum tolerated percentage of
  unclassified cases, $\gamma$, of~$.2$ for the four genes indicated
  at the top of each panel. 
The thinner line corresponds to the standard ROC curve. \label{fig:other}}
\end{figure}

Lastly, Figure~\ref{fig:other} shows fuzzy ROCs for four additional genes, chosen
in part to illustrate less common features. Regions can be disjoint, when
stretches of non-empty gray areas 
are followed by stretches of empty gray areas. Often this is
associated with lack of monotonicity in the likelihood ratio of the
two conditional biomarker distributions.

ZNF487 exemplifies a biomarker with relatively good
discrimination. The upper bounds indicates that correct
reclassification of as few as 20\% of cases could lead to high
discrimination. This reclassification could be achieved by biomarkers
that prove effective in the gray zone for ZNF487. The lower bound
indicates that, if unclassified observations are handled poorly, the
performance suffers, but discrimination remains above chance by a
clear margin even with a gray area of 20\%.

\section{Discussion}

I am not aware of a good visualization approach to examine
classification algorithms that allow for an area of indeterminacy. I
hope the fuzzy ROC approach will prove of practical help.

Allowing for a gray area differs from multi-class ROC analysis (e.g.\
\cite{Hand2001}), where the number of labels is greater than two. It
also differs from semisupervised analyses, where some cases are not
labeled. Here all cases have a known binary label but some are not classified. 

The fuzzyROC is not a visualization of uncertainty about the ROC curve
in the standard statistical sense. Both the upper and lower bound are
themselves point estimates, and their variability could be address by
simple resampling approaches. Yet visualizing both the set and
uncertainty about the 
set boundaries could be challenging. Also, $\gamma$ is expressed in terms of the (potentially rescaled) proportion of cases in the validation study, without consideration for uncertainty. 

The fuzzyROC is not an approach for optimizing the size $\gamma$ of the indeterminacy zone.
It only uses optimization to home in on a useful subset of options for visualization. 

The oracle and saboteur scenarios are extreme. Variants of this
algorithm could be constructed by further specifying bounds on the
proportion of cases that could be correctly classify by a human if
left in the gray area. Then instead of moving all the gray area points
to extremes, these known proportions could be used to move only some of the
points and achieve less extreme bounds. These classification
proportion could potentially depend on the biomarker region.

From a statistical perspective, indeterminacy can also help
characterize regions of the score with poor discriminatory
ability. Thus, compared to fully deterministic approaches, allowing
for indeterminacy may lead to a different evaluation of classifiers
and different approaches to biomarker discovery.


\section*{Acknowledgments}
Work supported by NIH grant 4P30CA006516-51 and NSF grant DMS-1810829.
Work currently submitted to Pattern Recognition Letters. 
The {\tt fuzzyROC} R package used to produce the analysis presented in this paper is freely available
for direct install at {\tt https://github.com/gp1d/fuzzyROC.git}.

%\section*{References}

\bibliographystyle{elsarticle-harv} 
\bibliography{fuzzyROC}

\begin{thebibliography}{6}
\expandafter\ifx\csname natexlab\endcsname\relax\def\natexlab#1{#1}\fi
\expandafter\ifx\csname url\endcsname\relax
  \def\url#1{\texttt{#1}}\fi
\expandafter\ifx\csname urlprefix\endcsname\relax\def\urlprefix{URL }\fi

\bibitem[{Fawcett(2006)}]{FAWCETT2006861}
Fawcett, T., 2006. An introduction to {ROC} analysis. Pattern Recognition
  Letters 27~(8), 861 -- 874, {ROC} Analysis in Pattern Recognition.
\newline\urlprefix\url{http://www.sciencedirect.com/science/article/pii/S016786550500303X}

\bibitem[{Ganzfried et~al.(2013)Ganzfried, Riester, Haibe-Kains, Risch,
  Tyekucheva, Jazic, Wang, Ahmadifar, Birrer, Parmigiani, Huttenhower, and
  Waldron}]{Ganzfried2013}
Ganzfried, B.~F., Riester, M., Haibe-Kains, B., Risch, T., Tyekucheva, S.,
  Jazic, I., Wang, X.~V., Ahmadifar, M., Birrer, M.~J., Parmigiani, G.,
  Huttenhower, C., Waldron, L., 2013. curated{O}varian{D}ata: clinically
  annotated data for the ovarian cancer transcriptome. Database (Oxford) 2013,
  bat013, pMCID: PMC3625954.
\newline\urlprefix\url{http://dx.doi.org/10.1093/database/bat013}

\bibitem[{Hand and Till(2001)}]{Hand2001}
Hand, D.~J., Till, R.~J., Nov 2001. A simple generalisation of the area under
  the roc curve for multiple class classification problems. Machine Learning
  45~(2), 171--186.
\newline\urlprefix\url{https://doi.org/10.1023/A:1010920819831}

\bibitem[{Parker et~al.(2009)Parker, Mullins, Cheang, Leung, Voduc, Vickery,
  Davies, Fauron, He, Hu, Quackenbush, Stijleman, Palazzo, Marron, Nobel,
  Mardis, Nielsen, Ellis, Perou, and Bernard}]{Parker:2009gxa}
Parker, J.~S., Mullins, M., Cheang, M. C.~U., Leung, S., Voduc, D., Vickery,
  T., Davies, S., Fauron, C., He, X., Hu, Z., Quackenbush, J.~F., Stijleman,
  I.~J., Palazzo, J., Marron, J.~S., Nobel, A.~B., Mardis, E., Nielsen, T.~O.,
  Ellis, M.~J., Perou, C.~M., Bernard, P.~S., Mar. 2009. {Supervised Risk
  Predictor of Breast Cancer Based on Intrinsic Subtypes}. Journal of Clinical
  Oncology 27~(8), 1160--1167.

\bibitem[{Riester et~al.(2014)Riester, Wei, Waldron, Culhane, Trippa, Oliva,
  Kim, Michor, Huttenhower, Parmigiani, and Birrer}]{Riester2014}
Riester, M., Wei, W., Waldron, L., Culhane, A.~C., Trippa, L., Oliva, E., Kim,
  S.-H., Michor, F., Huttenhower, C., Parmigiani, G., Birrer, M.~J., Apr 2014.
  Risk prediction for late-stage ovarian cancer by meta-analysis of 1525
  patient samples. J Natl Cancer Inst.
\newline\urlprefix\url{http://dx.doi.org/10.1093/jnci/dju048}

\bibitem[{Yoshihara et~al.(2012)Yoshihara, Tsunoda, Shigemizu, Fujiwara, Hatae,
  Fujiwara, Masuzaki, Katabuchi, Kawakami, Okamoto, Nogawa, Matsumura, Udagawa,
  Saito, Itamochi, Takano, Miyagi, Sudo, Ushijima, Iwase, Seki, Terao, Enomoto,
  Mikami, Akazawa, Tsuda, Moriya, Tajima, Inoue, and Tanaka}]{Yoshihara1374}
Yoshihara, K., Tsunoda, T., Shigemizu, D., Fujiwara, H., Hatae, M., Fujiwara,
  H., Masuzaki, H., Katabuchi, H., Kawakami, Y., Okamoto, A., Nogawa, T.,
  Matsumura, N., Udagawa, Y., Saito, T., Itamochi, H., Takano, M., Miyagi, E.,
  Sudo, T., Ushijima, K., Iwase, H., Seki, H., Terao, Y., Enomoto, T., Mikami,
  M., Akazawa, K., Tsuda, H., Moriya, T., Tajima, A., Inoue, I., Tanaka, K.,
  2012. High-risk ovarian cancer based on 126-gene expression signature is
  uniquely characterized by downregulation of antigen presentation pathway.
  Clinical Cancer Research 18~(5), 1374--1385.
\newline\urlprefix\url{http://clincancerres.aacrjournals.org/content/18/5/1374}

\end{thebibliography}

\end{document}